\documentclass[10pt,final,journal]{IEEEtran}

\usepackage{color,array,amsthm,amsmath,amsfonts}
\usepackage{graphicx}
\usepackage{cite}
\usepackage{graphicx}
\usepackage{textcomp}
\usepackage{xcolor}
\usepackage{verbatim}
\usepackage{array,multirow}
\usepackage{gensymb}

\begin{document}

\title{Toward High Accuracy DME for Alternative Aircraft Positioning: SFOL Pulse Transmission in High-Power DME} 

\author{
Jongmin~Park,
Seowoo~Park,
Sunghwa~Lee, 
Jiwon~Seo,~\IEEEmembership{Member,~IEEE}, and 
Euiho~Kim% <-this % stops a space
\thanks{Manuscript received June 00, 2025; }
\thanks{This research was supported in part by the Korea Institute of Marine Science \& Technology Promotion (KIMST), funded by the Ministry of Oceans and Fisheries, Republic of Korea, under Grant RS-2024-00407003;
and in part by the Korean government (Korea Aerospace Administration, KASA), under Grant RS-2022-NR067078.}
\thanks{Authors' addresses: Jongmin~Park and Sunghwa~Lee are with the School of Integrated Technology, Yonsei University, Incheon 21983, Republic of Korea, E-mail: (jm97@yonsei.ac.kr; sunghwa.lee@yonsei.ac.kr); 
Seowoo~Park and Euiho~Kim are with the Department of Mechanical and System Design Engineering, Hongik University, Seoul 04055, Republic of Korea, E-mail: (suwoo3131@gmail.com; euihokim@hongik.ac.kr);
Jiwon~Seo is with the School of Integrated Technology, Yonsei University, Incheon 21983, Republic of Korea and also with the Department of Convergence IT Engineering, Pohang University of Science and Technology, Pohang 37673, Republic of Korea, E-mail: (jiwon.seo@yonsei.ac.kr). \it{(Corresponding authors: Jiwon~Seo; Euiho~Kim.)}}% <-this % stops a space
}

\markboth{IEEE Transactions on Aerospace and Electronic Systems}{}
\maketitle

\begin{abstract}
The Stretched-FrOnt-Leg (SFOL) pulse is an advanced distance measuring equipment (DME) pulse that offers superior ranging accuracy compared to conventional Gaussian pulses. Successful SFOL pulse transmission has been recently demonstrated from a commercial Gaussian pulse-based DME in low-power mode utilizing digital predistortion (DPD) techniques for power amplifiers. These adjustments were achieved through software modifications, enabling SFOL integration without replacing existing DME infrastructure. However, the SFOL pulse is designed to optimize ranging capabilities by leveraging the effective radiated power (ERP) and pulse shape parameters permitted within DME specifications. Consequently, it operates with narrow margins against these specifications, potentially leading to non-compliance when transmitted in high-power mode. This paper introduces strategies to enable a Gaussian pulse-based DME to transmit the SFOL pulse while adhering to DME specifications in high-power mode. The proposed strategies involve use of a variant of the SFOL pulse and DPD techniques utilizing truncated singular value decomposition, tailored for high-power DME operations. Test results, conducted on a testbed utilizing a commercial Gaussian pulse-based DME, demonstrate the effectiveness of these strategies, ensuring compliance with DME specifications in high-power mode with minimal performance loss. This study enables cost-effective integration of high-accuracy SFOL pulses into existing high-power DME systems, enhancing aircraft positioning precision while ensuring compliance with industry standards.
\end{abstract}


\section{INTRODUCTION}
G{\scshape lobal} navigation satellite systems (GNSS) are widely utilized as a core technology for navigation and positioning across various industries owing to their exceptional accuracy and reliability \cite{Chen24:LSTM, Budtho24:Ground, Causa21:Improving, Lee22:Optimal, Jia21:Ground, Sabatini17:Global}.
However, they are susceptible to signal disruptions caused by natural phenomena, such as solar storms, or by intentional interference, such as jamming \cite{Reda24:Deep, Park21:Single, Liang24:Performance, Luo24:Zak, Wang18:GNSS, Moon24:HELPS, Lee22:Urban, Kim22:First, Son18:Novel, Lee17:Monitoring}.
To address the vulnerability posed by GNSS outages, the U.S. Federal Aviation Administration (FAA) has actively pursued the development of alternative position, navigation, and timing (APNT) systems for aircraft. 
Among the options considered, the FAA selected Distance Measuring Equipment (DME) as a short-term APNT solution through approximately 2030 \cite{FAA16:Performance}. 

Since the announcement of the performance-based navigation (PBN) strategy in 2016, ongoing research and innovation have explored the potential of DME as a long-term APNT solution---some of which are discussed below. 
DME is also recognized globally as a viable GNSS backup, with approximately 5,000 stations in operation as of 2021 \cite{Espen24:Old,Ostroumov21:Modelling}. 
The FAA has continued to expand its network of DME ground stations and, as of 2024, has completed the addition of 123 new stations to enhance area navigation coverage at the busiest airports supporting air carrier operations. 
Given the absence of other promising APNT alternatives for aircraft navigation, DME is expected to remain in operational use well beyond 2030 \cite{Lawrence24:FAA}.

DME is a pulse-based ranging system consisting of a ground transponder and an airborne interrogator \cite{Liu16:Adaptive, Hirsch49:Pulse}. 
The system determines the slant range between the transponder and interrogator by exchanging pairs of pulses and measuring the time-of-flight of the interrogation and reply pulses. 
A conventional DME typically employs a Gaussian pulse, known for its spectral power efficiency; this pulse minimizes interference with adjacent DME channels \cite{Yan23:Novel, Kim17:Improving}. 
However, the Gaussian pulse is limited by its susceptibility to multipath interference, which can result in ranging errors exceeding 100 m \cite{Tripathi24:Power, Kim17:Improving}.
This limitation is a potential reason for considering DME as only a short-term APNT solution, as its subpar ranging accuracy is a significant concern \cite{Kim17:Improving}. 

Numerous efforts have been made to enhance the ranging accuracy of DME \cite{Kim16:Multipath, Du24:Performance, Tripathi24:Power, Dzunda23:Influence}.
For example, Lo \textit{et al.} \cite{Lo14:Containing} proposed several methods for mitigating multipath effects in DME. First, they recommended conducting airspace surveys to pinpoint areas with significant multipath effects and then suggested adjusting operational procedures, such as implementing altitude limits, to mitigate these effects.
However, this approach necessitates substantial resources for surveying and may require frequent updates owing to environmental changes. 
Second, simple averaging techniques may not always be effective, as their performance significantly depends on the geometry between the DME ground station and aircraft. 
Third, although carrier smoothing and extended averaging can effectively suppress multipath effects when DME/N clock oscillators are stable, most DME equipment lacks the necessary stability. 
Li \textit{et al.} \cite{Li13:Enhanced} developed a method that utilizes the DME carrier as a ranging source, achieving centimeter-level accuracy. 
However, this approach entails replacing oscillators in existing Gaussian pulse-based DMEs, rendering it impractical for widespread implementation. 

\begin{figure} 
  \centering
  \includegraphics[width=0.9\linewidth]{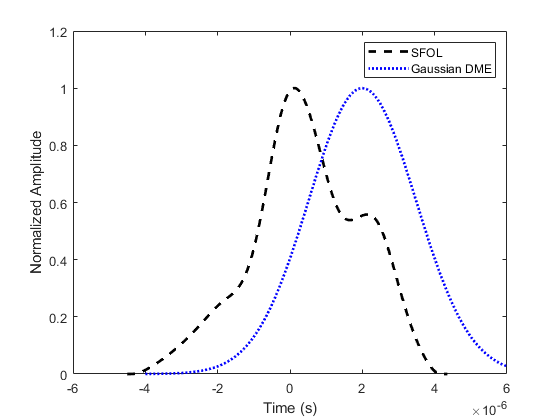}
  \caption{Comparison between the SFOL and Gaussian pulses.}
  \label{fig:ori_SFOL}
\end{figure}

Alternatively, Kim \cite{Kim13:Alternative, Kim17:Improving} and Kim and Seo \cite{Kim17:SFOL} proposed different types of DME pulses that suppress multipath effects, thereby enhancing ranging accuracy. 
These alternative pulses were designed to adhere to current DME specifications, ensuring compatibility with existing Gaussian pulse-based interrogators and transponders. 
Among these proposed pulses, the Stretched-FrOnt-Leg (SFOL) pulse demonstrated the best multipath mitigation performance, with a multipath-induced range error approximately 4.5 times smaller than that of a Gaussian pulse (Fig.~\ref{fig:ori_SFOL}) \cite{Kim17:SFOL}. 
A modern commercial Gaussian pulse-based DME can transmit an SFOL pulse with minimal modifications, maintaining compatibility with legacy DME systems.
If the SFOL pulse can be transmitted using only software updates, a high-accuracy DME system can be realized with minimal disruption.

\begin{figure*}
  \centering
  \includegraphics[width=0.7\linewidth]{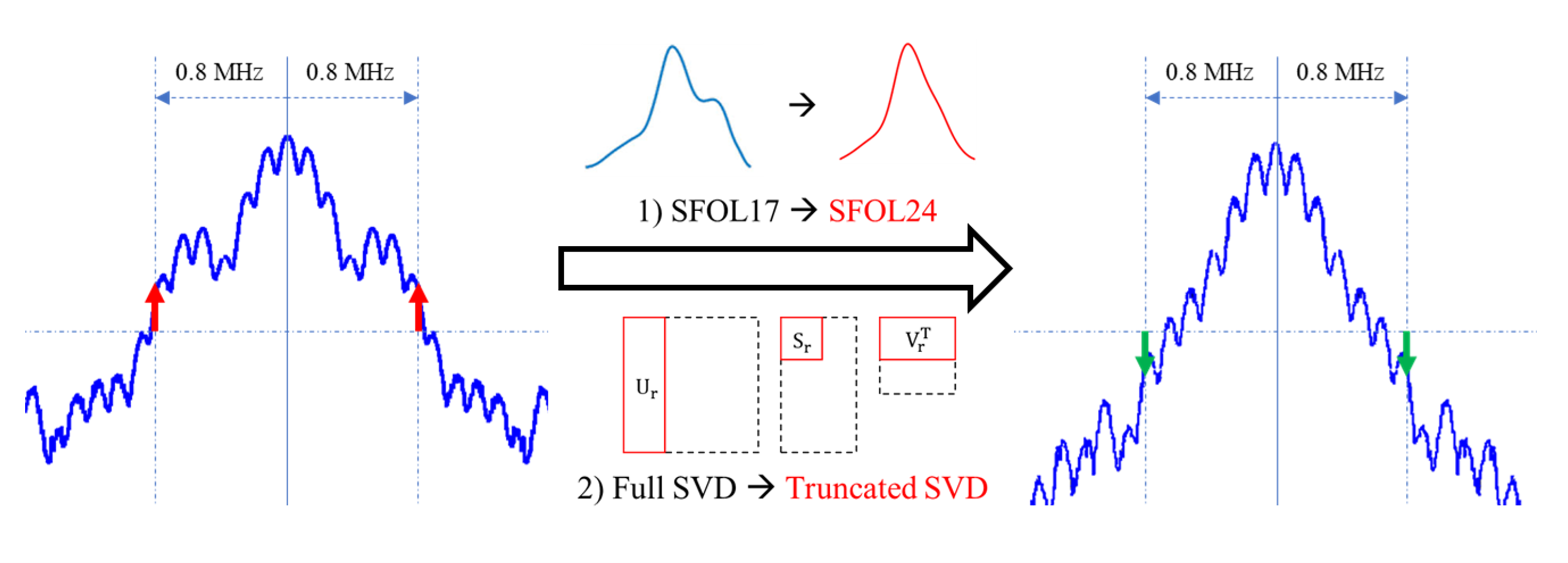}
  \caption{Overview of the challenges and proposed strategies for transmitting SFOL pulses in high-power DMEs.}
  \label{fig:contribution}
\end{figure*}

In our previous research, we conducted experiments to test the feasibility of transmitting SFOL pulses using a modern commercial Gaussian pulse-based DME by replacing the baseline Gaussian pulse with an SFOL pulse.
However, significant distortion in the transmitted SFOL pulses prevented them from maintaining their original shape. The primary source of this distortion was identified as the series of class-C power amplifiers (PAs), which introduced nonlinearities during  signal amplification \cite{Terman64:Electronic}. 

A DME digital predistortion (DPD) technique based on inverse learning was developed to address these nonlinearities.
This technique successfully transmitted the SFOL pulse in a 100 W low-power DME PA with solely software modifications \cite{Lee22:SFOL}. 
However, when applied to a 1000 W high-power DME, the transmitted SFOL pulse exceeded DME power spectrum specifications.
This can be attributed to the narrow margins of the SFOL pulse against the DME spectral specifications, whereas a typical commercial DME has a large noise floor in the transmitter unit that is not effectively treated by a conventional DPD process.

To  overcome these challenges, this study proposes two methods: a modified version of the SFOL pulse tailored for high-power DMEs and the utilization of truncated singular value decomposition (TSVD) in the DPD process (Fig.~\ref{fig:contribution}).
To prevent any confusion, the SFOL pulse discussed in \cite{Kim17:SFOL} has been designated as the SFOL17 pulse, whereas the newly proposed variant is referred to as the SFOL24 pulse.
The SFOL24 pulse was generated using genetic algorithms with stricter spectral specifications, resulting in reduced spectral power while maintaining multipath ranging accuracy comparable with that of the SFOL17 pulse.
TSVD was implemented to counteract noise effects during the DPD process.
These combined strategies enable the SFOL24 pulse to be transmitted in 1000 W high-power DMEs with minimal performance degradation compared to the SFOL17 pulse.

The remainder of this paper is organized as follows.
Section \ref{sec:prev_studies} reviews previous research on DPD design for 100 W low-power DMEs, highlighting persistent challenges and introducing relevant DME specifications.
Section \ref{sec:strategies} proposes two strategies for transmitting the variant SFOL pulse in a 1000 W high-power DME.
Section \ref{sec:results} details the design of the DME transmitter testbed and evaluates the proposed strategies through experimental results.
Finally, Section \ref{sec:conclusion} presents concluding remarks.

\section{PREVIOUS STUDIES AND PERSISTENT CHALLENGES}
\label{sec:prev_studies}
\subsection{Transmission of SFOL17 Pulse using DPD} 

\begin{figure} 
  \centering
  \includegraphics[width=1\linewidth]{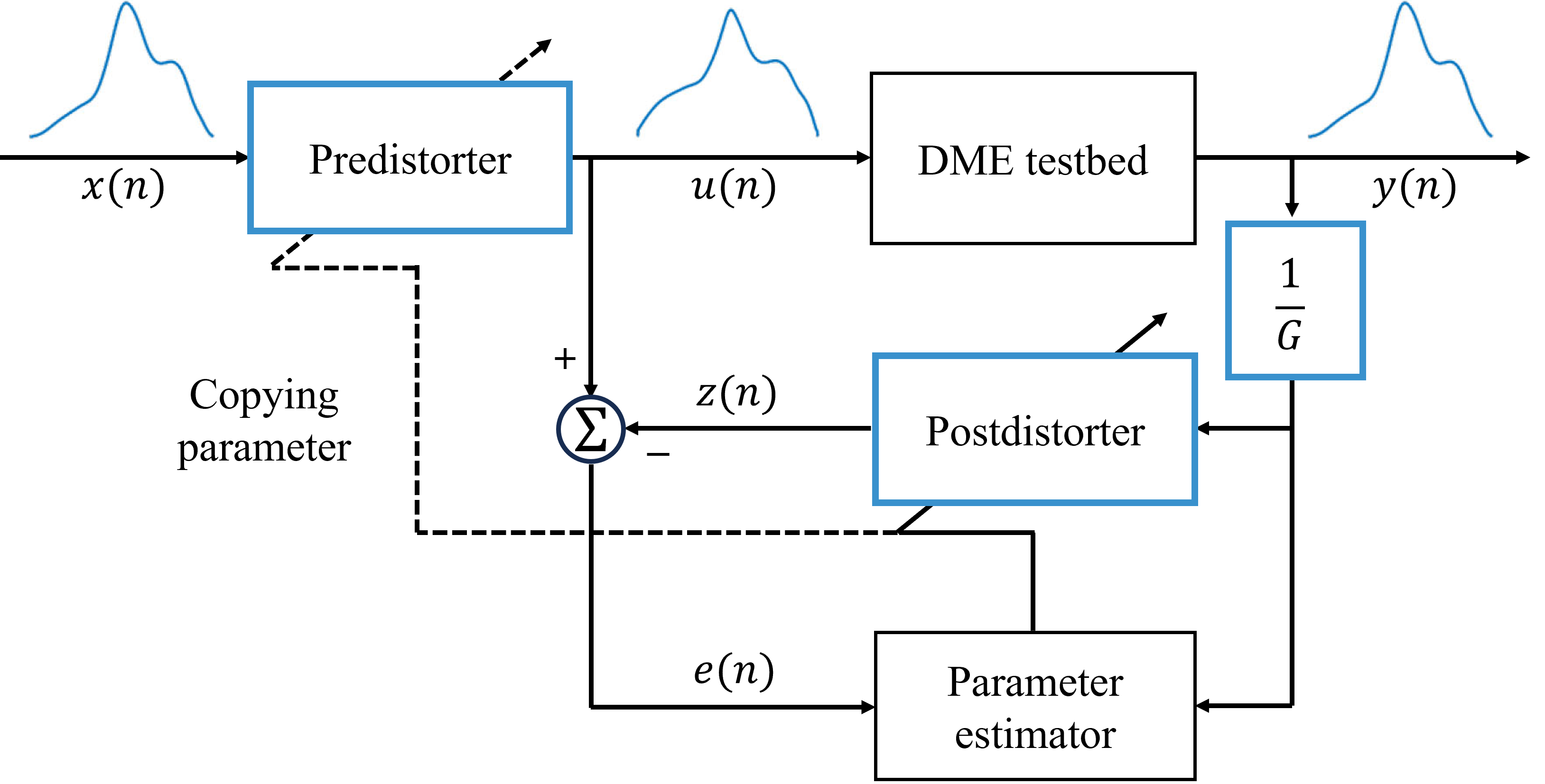}
  \caption{Block diagram of DPD for DME in a previous study  (reproduced from Fig. 1 of \cite{Lee22:SFOL}).}
  \label{fig:DPDdiag}
\end{figure}

\begin{figure*}
  \centering
  \includegraphics[width=0.7\linewidth]{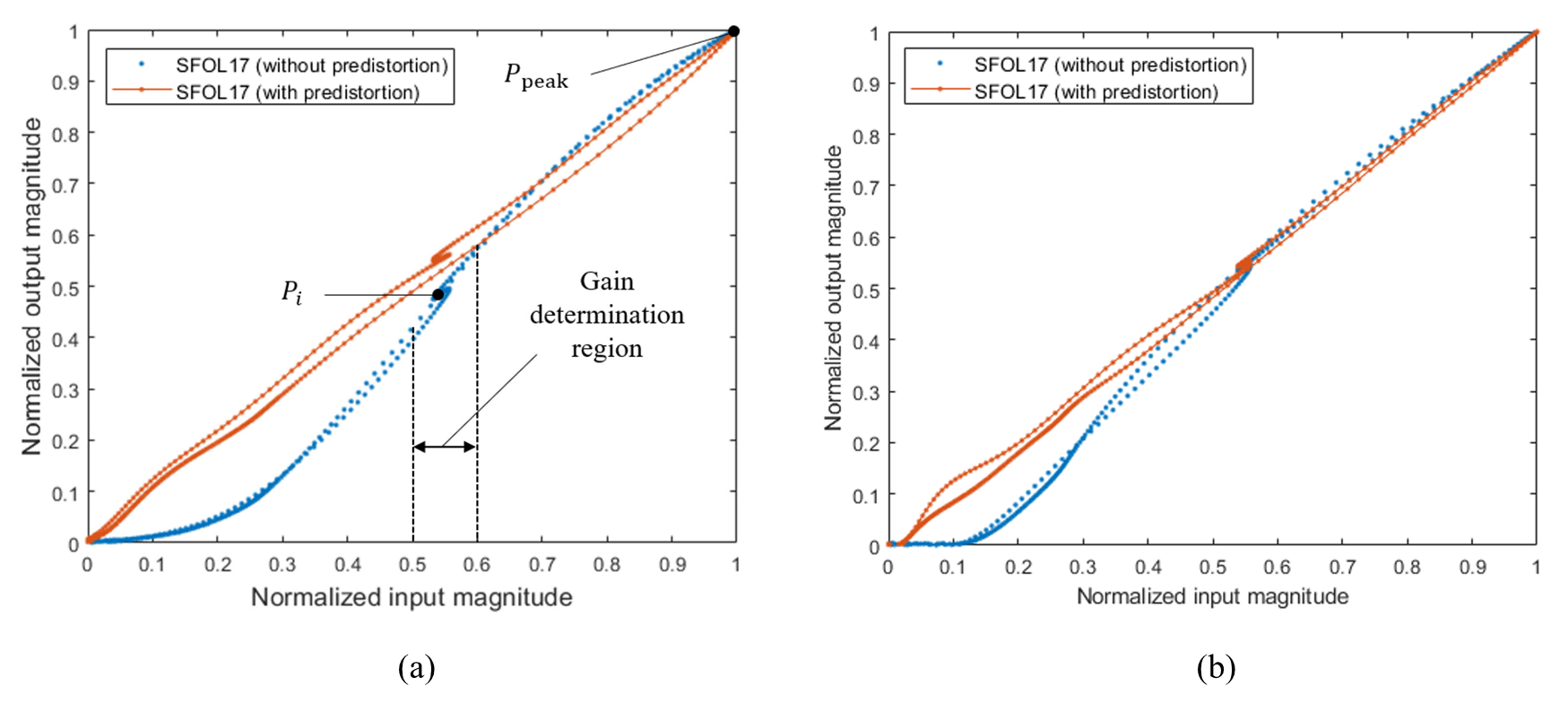}
  \caption{Normalized output versus input magnitudes in (a) low-power DME (reproduced from Fig. 2 of \cite{Lee22:SFOL}) and (b) high-power DME.}
  \label{fig:gaindet_combined}
\end{figure*}

As previously mentioned, the transmission of SFOL17 pulses in contemporary commercial Gaussian pulse-based DME systems has resulted in significant signal distortion owing to the nonlinearity induced by the PA. 
The Gaussian pulse, despite its distorted shape, causes minimal interference with adjacent DME channels in terms of spectral power, ensuring compliance with DME specifications.
In contrast, the SFOL17 pulse is close to the specification limits, necessitating compensation for signal distortion to satisfy DME standards. 

To address this issue, a DME DPD technique was introduced that successfully transmitted SFOL17 pulses using a 100 W low-power DME PA \cite{Lee22:SFOL}. 
The overall DPD procedure for DME outlined in \cite{Lee22:SFOL} is shown in Fig. \ref{fig:DPDdiag}.
An RF input digital waveform, $x(n)$, where $n$ represents the discrete time index indicating the sample position, is transformed into a predistorted waveform, $u(n)$, using a predistorter.
The waveform $u(n)$ is then distorted to $y(n)$ after it passes through multi-chain pulse-shaping circuits and PA modules.
The distorted waveform $y(n)$ is normalized by applying the gain $G$, and then passed to a postdistorter and  parameter estimator. 
This process is repeated to enable the predistorter to converge.

When the SFOL17 waveform is inserted into low-power DME hardware without predistortion, the normalized output magnitude is significantly low at low input levels, as shown in Fig.~\ref{fig:gaindet_combined}a.
Lee \textit{et al.} \cite{Lee22:SFOL} tailored a DPD technique for DME by modifying the gain normalization method and adding a bias term to the memory polynomial model.
In conventional DPD implementations for PAs, the normalization gain is typically set either to the maximum gain within the linear region or to the gain corresponding to the maximum output power, both of which yield similar performance \cite{Zhu08:Open}. 
In \cite{Lee22:SFOL}, the gain determination region was chosen based on high linearity, and the gain was calculated within this region (Fig.~\ref{fig:gaindet_combined}a).

The memory polynomial model is widely utilized to model PA owing to its effectiveness and low computational complexity \cite{Lei04:Robust, Morgan06:Generalized, Tafuri12:Linearization, Chen21:Modified}.
A bias term has been added to the memory polynomial model to effectively model an extremely low normalized output power at low input levels. 
These methods for DME were proven to outperform methods that use the gain at the maximum output power or a memory polynomial model without bias \cite{Lee22:SFOL}.

High-power DME exhibits consistently low normalized output power at low input power levels, mirroring the behavior of low-power DME without predistortion (Fig. ~\ref{fig:gaindet_combined}b).
%Specifically, the normalized output power remains below 0.01 until approximately 1.7 µs. 
To address this issue, we have implemented the gain normalization method and memory polynomial model with a bias, as proposed in a previous study.

\subsection{Test Results in Low-Power and High-Power DME Testbeds and Persistent Challenges}

The DME pulse must satisfy the following two key specifications. First, it must not surpass the DME power spectrum limits established by the International Civil Aviation Organization (ICAO) or the FAA. These specifications dictate that the maximum effective radiated power (ERP) within a 0.5 MHz band centered at 0.8 MHz and 2 MHz away from the center frequency should not exceed 23.0 dBm and 3.0 dBm, respectively, in 1000 W high-power DME \cite{ICAO23:Annex, FAA21:Performance}.

Second, the pulse shape must fall within the rise time, width, and fall time parameters outlined by ICAO and FAA \cite{ICAO23:Annex, FAA21:Performance}.
Rise time refers to the duration taken for a pulse to transition from 10\% to 90\% of its peak amplitude on the rising edge, whereas fall time represents the time taken for the pulse to transition from 90\% to 10\% of its peak amplitude on the falling edge. 
Width is defined as the time interval between the pulse reaching 50\% of its peak amplitude on the rising and falling edges. 
The specifications are detailed in Tables \ref{tab:Power_Spectrum_specifications} and \ref{tab:Pulse_Shape_specifications}, alongside the theoretical power spectra and pulse shape of SFOL17.

\begin{table}
\centering
\caption{Theoretical power spectra of SFOL17 compared with the DME power spectrum specifications \cite{Kim17:SFOL}.}
\label{tab:Power_Spectrum_specifications}
\renewcommand{\arraystretch}{1.2}
\begin{tabular}{ccc}
\hline
Adjacent channel (MHz)     & $\pm0.8$ & $\pm2.0$ \\ \hline
Specifications (dBm)           & 23.0            &   3.0           \\
SFOL17 (dBm)           & 22.0             & $-11.5$          \\ \hline
\end{tabular}
\end{table}

\begin{table}
\centering
\caption{Theoretical pulse shape of SFOL17 compared with ICAO and FAA DME pulse shape specifications \cite{Kim17:SFOL}.}
\label{tab:Pulse_Shape_specifications}
\renewcommand{\arraystretch}{1.2}
\resizebox{\columnwidth}{!}{
\begin{tabular}{crrr}
\hline
              & Rise time ($\mathrm{\mu s})$         & Width ($\mathrm{\mu s}$)          & Fall time ($\mathrm{\mu s})$        \\ \hline
ICAO specifications & $< 3.0$  & 3.5($\pm0.5$)  & 3.0($+0.5$,$-0.5$)  \\
FAA specifications  & 2.5($+0.5$,$-1.0$)  & 3.5($\pm0.5$)  & 2.5($+0.5$,$-1.0$)      \\
SFOL17 & 2.79          & 3.43        & 3.00          \\ \hline
\end{tabular}
}
\end{table}

The DPD technique successfully transmitted the SFOL17 pulse through a 100 W low-power DME PA with only software changes \cite{Lee22:SFOL}. 
The previous test results from \cite{Lee22:SFOL}, obtained using the DPD technique on the low-power DME testbed, are listed in Tables~\ref{tab:LPA_Power_Spectrum_Results} and \ref{tab:LPA_Pulse_Shape_Results}. 
A 10 dB adjustment was made to the low-power DME test results for comparison with the spectrum specifications given for the 1000 W high-power DME, as the 1000 W high-power DME operates 10 dB above the 100 W low-power DME. 
As indicated in Tables \ref{tab:LPA_Power_Spectrum_Results} and \ref{tab:LPA_Pulse_Shape_Results}, the previous DPD technique satisfied both the ICAO DME pulse shape specifications and DME power spectrum specifications for the low-power DME testbed.
(However, it did not satisfy the FAA DME pulse shape specifications.)

\begin{table} 
\centering
\caption{Low-power DME power spectra compared with DME power spectrum specifications when SFOL17 and the previous DPD technique were applied (test results from \cite{Lee22:SFOL}).}
\label{tab:LPA_Power_Spectrum_Results}
\renewcommand{\arraystretch}{1.2}
\resizebox{\columnwidth}{!}{
\begin{tabular}{ccccc}
\hline
Adjacent channel (MHz) & $-2.0$ & $-0.8$ & $+0.8$ & $+2.0$ \\ \hline
Specifications (dBm) & 3.0 & 23.0   & 23.0  &   3.0             \\
Low-power DME $+$ 10 dB (dBm) & $-2.9$  & 19.1  & 20.6 & $-3.6$            \\ \hline
\end{tabular}
}
\end{table}

\begin{table}
\centering
\caption{Low-power DME pulse shapes compared with ICAO and FAA DME pulse shape specifications when SFOL17 and the previous DPD technique were applied (test results from \cite{Lee22:SFOL}).}
\label{tab:LPA_Pulse_Shape_Results}
\renewcommand{\arraystretch}{1.2}
\resizebox{\columnwidth}{!}{
\begin{tabular}{crrr}
\hline
              & Rise time ($\mathrm{\mu s}$)        & Width ($\mathrm{\mu s}$)         & Fall time ($\mathrm{\mu s}$)       \\ \hline
ICAO specifications & $< 3.0$ & 3.5($\pm0.5$) & 3.0($+0.5$,$-0.5$)  \\
FAA specifications  & 2.5($+0.5$,$-1.0$) & 3.5($\pm0.5$) & 2.5($+0.5$,$-1.0$)  \\
Low-power DME & 2.88 & 3.46   & 3.05  \\ \hline
\end{tabular}
}
\end{table}

Utilizing the DPD technique in \cite{{Lee22:SFOL}}, we conducted tests to determine whether the SFOL17 pulses could be successfully transmitted on a high-power DME testbed. 
The findings are presented in Tables \ref{tab:HPA_Power_Spectrum_specifications} and \ref{tab:HPA_Pulse_Shape_Results}. 
The data in these tables demonstrate that the previous DPD technique failed to satisfy the DME power spectrum specifications for the high-power DME.

\begin{table} 
\centering
\caption{High-power DME power spectra compared with DME power spectrum specifications when SFOL17 and the previous DPD technique were applied.}
\label{tab:HPA_Power_Spectrum_specifications}
\renewcommand{\arraystretch}{1.2}
\begin{tabular}{ccccc}
\hline
Adjacent Channel (MHz) & $-2.0$ & $-0.8$ & $+0.8$ & $+2.0$\\ \hline
Specifications (dBm) & 3.0 & 23.0  & 23.0   & 3.0            \\
High-power DME (dBm) & 7.7  & 24.9 & 25.4 & 6.5            \\ \hline
\end{tabular}
\end{table}

\begin{table} 
\centering
\scriptsize
\caption{High-power DME pulse shapes compared with ICAO and FAA DME pulse shape specifications when SFOL17 and the previous DPD technique were applied.}
\label{tab:HPA_Pulse_Shape_Results}
\renewcommand{\arraystretch}{1.2}
\resizebox{\columnwidth}{!}{
\begin{tabular}{crrr}
\hline
              & Rise time ($\mathrm{\mu s}$)   & Width ($\mathrm{\mu s}$) & Fall time ($\mathrm{\mu s}$)       \\ \hline
ICAO specifications & $< 3.0$  & 3.5($\pm0.5$) & 3.0($+0.5$,$-0.5$)  \\
FAA specifications  & 2.5($+0.5$,$-1.0$) & 3.5($\pm0.5$) & 2.5($+0.5$,$-1.0$)      \\
High-power DME & 2.78          & 3.34      & 3.01          \\ \hline
\end{tabular}
}
\end{table}

We hypothesized that this failure (i.e., the ERP of the high-power DME exceeding the DME power spectrum specifications) is due to the significant noise floor in the transmitter unit. 
To validate this assumption, we conducted experiments to obtain 100 transmitted SFOL17 pulses without predistortion from both low- and high-power DMEs.
Subsequently, the noise-to-signal ratios (NSR) of the low-power and high-power DMEs, denoted as $NSR_{\mathrm{LP}}(n)$ and $NSR_{\mathrm{HP}}(n)$, were calculated as follows:
\begin{equation}
    NSR_{\mathrm{LP}}(n) = \frac{{\sigma}_{\mathrm{LP}}(n)}{\bar{y}_{\mathrm{LP}}(n)}, \hspace{0.5cm} NSR_{\mathrm{HP}}(n) = \frac{{\sigma}_{\mathrm{HP}}(n)}{\bar{y}_{\mathrm{HP}}(n)}
    \label{eqn:NSR}
\end{equation}
where $\bar{y}_{\mathrm{LP}}(n)$ and $\bar{y}_{\mathrm{HP}}(n)$ denote the mean of the 100 transmitted pulses $y(n)$ at each sample point $n$ from low- and high-power DMEs, respectively. 
Similarly, $\sigma_{\mathrm{LP}}(n)$ and $\sigma_{\mathrm{HP}}(n)$ denote the standard deviation of the 100 transmitted pulses $y(n)$ at each sample point $n$ from low- and high-power DMEs, respectively. 

\begin{figure}
  \centering
  \includegraphics[width=0.9\linewidth]{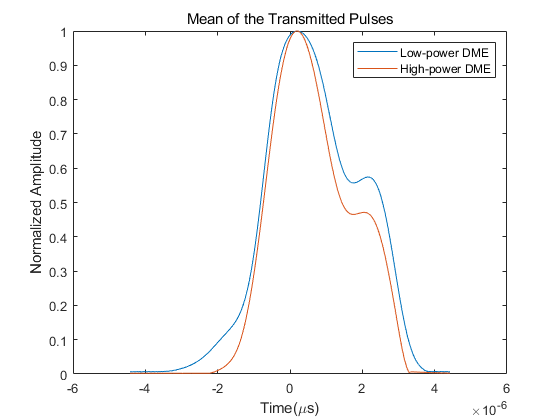}
  \caption{Mean of the transmitted SFOL17 pulses from low- and high-power DMEs.}
  \label{fig:mean_pulse}
\end{figure}

\begin{figure}
  \centering
  \includegraphics[width=0.9\linewidth]{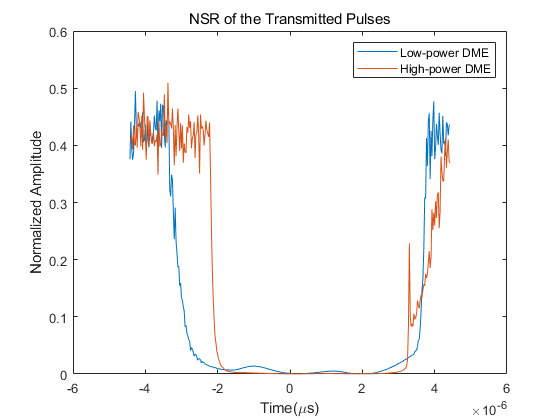}
  \caption{NSR of the transmitted SFOL17 pulses from low- and high-power DMEs.}
  \label{fig:NSR}
\end{figure}

The experimental results of the analysis of the transmitted SFOL17 pulses from the low- and high-power DMEs are shown in Figs. \ref{fig:mean_pulse} and \ref{fig:NSR}. 
The mean values of the transmitted pulses, denoted as $\bar{y}_{\mathrm{LP}}(n)$ and $\bar{y}_{\mathrm{HP}}(n)$ are shown in Fig.~\ref{fig:mean_pulse}, while a comparison of the NSR values of the transmitted pulses, denoted as $NSR_{\mathrm{LP}}(n)$ and $NSR_{\mathrm{HP}}(n)$, is shown in Fig.~\ref{fig:NSR}. 
The total NSR for the high-power DME, $\sum_{n=1}^{N}{NSR_{\mathrm{HP}}(n)}$, where $N = 500$ represents the total number of sample points, was calculated as 49.70, exceeding that of the low-power DME, $\sum_{n=1}^{N}{NSR_{\mathrm{LP}}(n)}$, which was calculated as 38.07. 
This result indicates that a more noise-robust DPD technique is required for high-power DMEs compared to low-power DMEs.
The memory polynomial model, due to its near-singular matrix, is sensitive to noise in $y(n)$, necessitating a solution to this issue \cite{Gilabert20:Beyond, Hansen87:truncatedSVD, Chen19:Toward}. 
To resolve this issue, we enhanced the DPD technique in this study by adapting TSVD, a method known for its robustness against noise \cite{Hansen87:truncatedSVD, Chen19:Toward, Balatti21:Aircraft}.

\section{STRATEGIES FOR TRANSMITTING SFOL PULSES IN HIGH-POWER DMES}
\label{sec:strategies}

It is observed in Table \ref{tab:HPA_Power_Spectrum_specifications} that the ERP of a transmitted SFOL17 pulse using the previous DPD technique in a high-power DME surpassed the ERP thresholds outlined in the specifications. 
The high ERP is attributed to a significant noise floor and imperfect signal processing within the transmitter unit of modern DMEs utilizing Gaussian pulses. 
These issues, particularly hardware-related, proved challenging to rectify solely through conventional DPD techniques. 
Notably, the aforementioned issues are not problematic in Gaussian pulse-based DMEs, as the ERP of Gaussian pulses is sufficiently low, preventing the combined effects of noise and distortion from exceeding ERP thresholds, even in high-power DMEs.

In contrast, the ERP of SFOL17 pulses is much closer to the ERP threshold. 
Specifically, at a frequency offset of 0.8 MHz from the center frequency, the ERP threshold is 23.0 dBm, whereas the theoretical ERP of the SFOL17 pulse is 22.0 dBm (Table~\ref{tab:Power_Spectrum_specifications}).
This narrow margin suggests that even slight distortions between the transmitted and input pulses could result in the ERP exceeding the threshold.
To mitigate this risk, it is essential to partially modify the SFOL17 pulse and implement noise-robust DPD techniques.
This section outlines viable solutions to enable SFOL-like pulse transmissions in high-power DMEs.

\subsection{Variant SFOL Pulse through Tightened Pulse Specifications}

\begin{figure} 
  \centering
  \includegraphics[width=0.8\linewidth]{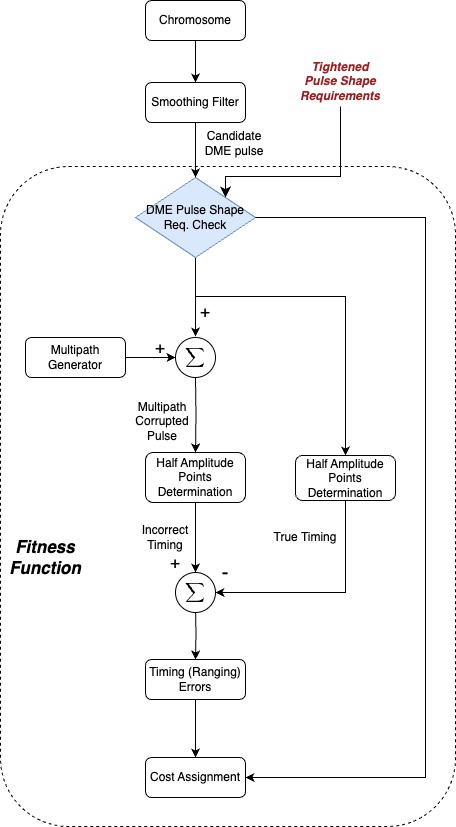}
  \caption{Overall process of the genetic algorithm used to develop the SFOL17 pulse (adapted from Fig. 4 in \cite{Kim17:SFOL}).}
  \label{fig:Fitness}
\end{figure}

The SFOL17 pulse was initially developed using genetic algorithms \cite{Kim17:SFOL}.
In this approach, a DME pulse was modeled as a chromosome comprising 64 genes, each ranging from 0 to 1 and corresponding to an amplitude point of a pulse over 12 $\mu$s period.
Subsequently, a smooth DME pulse was generated by interpolating these amplitude samples in a chromosome using a piece-wise cubic spline function. 
The process for identifying an optimal chromosome for a DME pulse is shown in Fig.~\ref{fig:Fitness}.

The SFOL17 pulse utilizes most of the available DME power spectrum; hence, designing an SFOL-like pulse with a lower ERP while maintaining competitive multipath mitigation performance would prove advantageous. 
To design a new variant SFOL (SFOL24) pulse, we employed genetic algorithms with two key modifications in the initial pulse shape and fitness function. 
First, the initial chromosome or pulse shape was changed from a Gaussian to the SFOL17 pulse, which resulted in an SFOL24 pulse with a shape similar to that of the SFOL17 pulse. 
Additionally, the fitness function was adjusted to tighten the maximum allowable ERPs from 23.0 to 16.0 dBm at 0.8 MHz from the DME channel frequency.

\begin{figure} 
  \centering
  \includegraphics[width=1\linewidth]{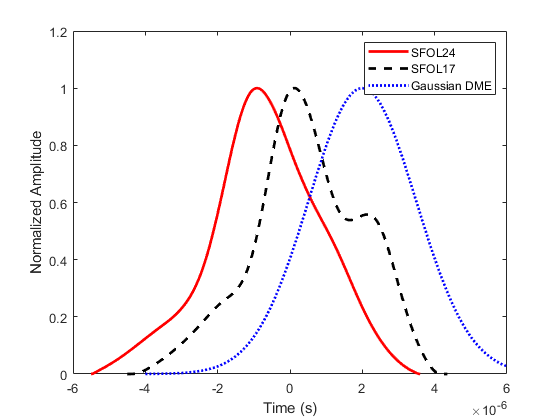}
  \caption{Comparison of SFOL17 and SFOL24 pulse shapes.}
  \label{fig:SFOL24}
\end{figure}

The shapes of the SFOL17, SFOL24, and Gaussian pulses are shown in Fig.~\ref{fig:SFOL24}. 
The SFOL24 pulse exhibits a considerably smoother profile compared to the SFOL17 pulse, particularly at its falling edge, whereas its rising edge closely resembles that of the SFOL17 pulse.
This characteristic is crucial, as the gradual rise followed by a rapid slope, as seen in the SFOL17 pulse, is crucial in mitigating multipath-induced range errors. 
The rise time, width, and fall time of the SFOL24 pulse are listed in Table~\ref{tab:Pulse_Shape_specifications_PMSFOL}.

\begin{table}
\centering
\caption{Comparison of theoretical SFOL17 and SFOL24 pulse shapes with ICAO and FAA DME pulse shape specifications.}
\label{tab:Pulse_Shape_specifications_PMSFOL}
\renewcommand{\arraystretch}{1.2}
\resizebox{\columnwidth}{!}{
\begin{tabular}{ccccc}
\hline
                       & Rise time ($\mathrm{\mu s}$)        & Width ($\mathrm{\mu s}$)         & Fall time ($\mathrm{\mu s}$)       \\ \hline
ICAO specifications & $< 3.0$ & 3.5($\pm0.5$)  & 3.0($0.5$,$-0.5$) \\
FAA specifications  & 2.5($+0.5$,$-1.0$)  & 3.5($\pm0.5$)  & 2.5($+0.5$,$-1.0$)      \\
SFOL17           & 2.79  & 3.43        & 3.00        \\
SFOL24 & 2.87       & 3.14      & 2.95         \\ \hline
\end{tabular}
}
\end{table}

\begin{figure}
  \centering
  \includegraphics[width=1\linewidth]{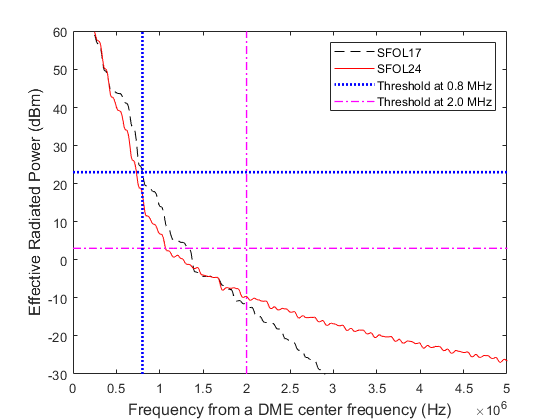}
  \caption{Comparison of the ideal ERPs of undistorted SFOL17 and SFOL24 pulses.}
  \label{fig:ERP_SFOL24}
\end{figure}

The theoretical power spectra at 0.8 MHz and 2 MHz away from the center frequency are listed in Table~\ref{tab:Power_Spectrum_specifications_PMSFOL}.
Notably, the ERP of the SFOL24 pulse decreases to 15.5 dBm at 0.8 MHz away from the center frequency, whereas at 2 MHz, the SFOL24 pulse's ERP slightly exceeds that of the SFOL17 pulse. 
However, the ERP level of the SFOL24 pulse at 2 MHz away from the center frequency remains within acceptable limits, well below the established threshold.
A comparison of the overall ERPs of the SFOL17 and SFOL24 pulses is shown in Fig.~\ref{fig:ERP_SFOL24}.

\begin{table}
\centering
\caption{Comparison of the theoretical power spectra of SFOL17 and SFOL24 pulses with ICAO and FAA DME power spectrum specifications.}
\label{tab:Power_Spectrum_specifications_PMSFOL}
\renewcommand{\arraystretch}{1.2}
\begin{tabular}{ccc}
\hline
Adjacent channel (MHz)     & $\pm0.8$ & $\pm2.0$ \\ \hline
Specifications (dBm)           & 23.0            &   3.0           \\
SFOL17 (dBm)           & 22.0             & $-11.5$          \\
SFOL24 (dBm)           & 15.5           &  $-11.3$          \\ \hline
\end{tabular}
\end{table}

The multipath mitigation performance was evaluated by calculating the multipath-induced range error using the method outlined in the previous study \cite{Kim17:SFOL}.
The DME multipath $m(t)$ is modeled as follows:
 \begin{equation}
  m(t,\phi,\delta)=\alpha\cos(\phi)y(t-\delta)
  \label{eqn:multi}
\end{equation}
where $t$ denotes the sampling time, $y(t)$ represents the direct pulse, $\alpha$ represents the peak amplitude ratio of the multipath to the direct pulse (set to 0.3), $\phi$ is the phase difference between the direct and multipath signals (ranging from 0 to $2\pi$), and $\delta$ represents the delay time of the multipath relative to the direct pulse (ranging from 0 to 6 $\mathrm{\mu s}$). 
The rationale behind using this configuration is that any delay in the multipath signal exceeding 6 $\mathrm{\mu s}$ would have minimal impact on the rising edge of the direct pulse.

\begin{figure}
  \centering
  \includegraphics[width=1\linewidth]{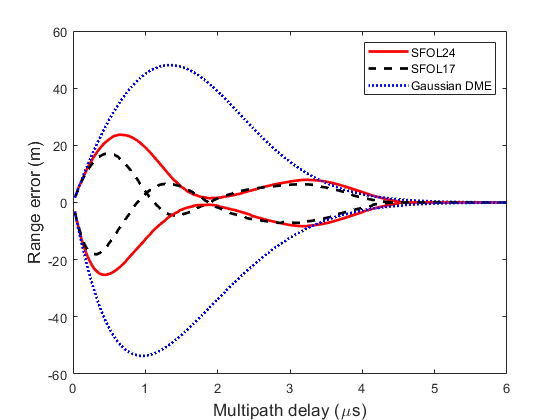}
  \caption{Comparison of the multipath-induced range errors of the SFOL17 and SFOL24 pulses.}
  \label{fig:SFOL24_mp}
\end{figure}

The multipath-induced range errors of the Gaussian, SFOL17, and SFOL24 pulses, with a multipath to direct pulse peak voltage ratio of 0.3, are shown in Fig.~\ref{fig:SFOL24_mp} and Table~\ref{tab:MP_ranges}.
The largest multipath-induced range error for the SFOL24 pulse is $-25.3$ m, whereas that for the SFOL17 pulse is $-18.1$ m. 
The overall RMS of the multipath-induced range errors for the SFOL17 and SFOL24 pulses are 6.0 m and 9.2 m, respectively.
Compared to the multipath-induced range errors of the Gaussian pulse, the SFOL24 pulse demonstrates a strong multipath resistance capability, similar to that of the SFOL17 pulse.

\begin{table}
\centering
\caption{Largest multipath-induced range errors of the three pulses in both in-phase and out-of-phase conditions, as well as the RMS of the entire range errors.}
\label{tab:MP_ranges}
\renewcommand{\arraystretch}{1.2}
\begin{tabular}{cccc}
\hline
                        & In-phase (m) & Out-of-phase (m) & RMS (m) \\ \hline
Gaussian           & 48.1              &   $-53.7$  &    26.1         \\
SFOL17           & 17.1              & $-18.1$     & 6.0         \\
SFOL24                  & 23.8             &  $-25.3$    & 9.2          \\ \hline
\end{tabular}
\flushleft
\noindent{\footnotesize Note: The direct-to-multipath peak voltage ratio is 1:0.3.}
\end{table}

\subsection{Enhanced DPD using TSVD} 

In the DPD process shown in Fig.~\ref{fig:DPDdiag}, the power amplifier (PA) output $y(n)$ is used to determine the predistorter and postdistorter parameters. 
A memory polynomial PA model is employed, where the PA output sample $y(n)$ from the input signal sample $u(n)$ is expressed as follows:
\begin{equation}
  y(n) = \sum_{k=0}^{K-1} \sum_{m=0}^{M-1} a_{km} u(n-m) {\lvert u(n-m) \rvert}^k
  \label{eqn:yMP}
\end{equation}
where $K$ and $M$ denote the nonlinearity order and memory depth, respectively, and $a_{km}$ represents the model coefficients. 
A previous study \cite{Lee22:SFOL} incorporated a bias term in the memory polynomial model, demonstrating its effectiveness. 
With this modification, the PA output sample $y(n)$ is expressed as:
\begin{equation}
  y(n) = \sum_{k=0}^{K-1} \sum_{m=0}^{M-1} a_{km} u(n-m) {\lvert u(n-m) \rvert}^k+b.
  \label{eqn:yMPb}
\end{equation}

With a set of signal samples $\mathbf{y}= [ y(1), \dots, y(L)]^\top$ and $\mathbf{u}= [ u(1), \dots, u(L)]^\top$, (\ref{eqn:yMPb}) can be rewritten as: 
\begin{equation}
\mathbf{y}_{L \times 1} = \mathbf{U}_{L \times (KM+1)} \mathbf{a}_{(KM+1) \times 1}
\label{eq:yua}
\end{equation}
where
\begin{equation*}
\mathbf{U}_{L \times (KM+1)}=\left[\begin{array}{llllll}
\mathbf{U}_{0} & \mathbf{U}_{1} & \mathbf{U}_{2} & \dots & \mathbf{U}_{M-1} & \mathbf{1}
\end{array}\right],
\end{equation*}

\begin{frame}
\footnotesize
\setlength{\arraycolsep}{2.5pt} % default: 5pt
\medmuskip = 1mu % default: 4mu plus 2mu minus 4mu
\begin{equation*}
    \mathbf{U}_{m}=
    \begin{bmatrix}
u{(1-m)} &  \dots & u{(1-m)}\left|u{(1-m)}\right|^{K-1}  \\
u{(2-m)} &  \dots & u{(2-m)}\left|u{(2-m)}\right|^{K-1}  \\
u{(3-m)} &  \dots & u{(3-m)}\left|u{(3-m)}\right|^{K-1} \\
\vdots & \ddots & \vdots \\
u{(L-m)} & \dots & u{(L-m)}\left|u{(L-m)}\right|^{K-1} 
\end{bmatrix}_{L \times K},
\end{equation*}
\end{frame}
\begin{equation*} 
\mathbf{1}_{L \times 1}=\left[\begin{array}{llllll}
1 & 1 & 1 & \dots & 1
\end{array}\right]^\top, 
\end{equation*}
and 
\begin{equation*} 
\mathbf{a}_{(KM+1) \times 1}=\left[\begin{array}{llllll}
{a}_{00} & {a}_{10} & {a}_{20} & \dots & {a}_{(K-1)(M-1)} & b
\end{array}\right]^\top. 
\end{equation*}

The postdistorted signal, $z(n)$, is expressed as:
 \begin{equation}
  z(n) = \frac{1}{G}\sum_{k=0}^{K-1} \sum_{m=0}^{M-1} a_{km} y(n-m) {\lvert y(n-m) \rvert}^k+b
  \label{eqn:zMP}
\end{equation}
and the predistorted signal, $u(n)$, is expressed as:
\begin{equation}
  u(n) = \sum_{k=0}^{K-1} \sum_{m=0}^{M-1} a_{km} x(n-m) {\lvert x(n-m) \rvert}^k+b.
\end{equation} 
Similar to (\ref{eq:yua}), the two signals can be rewritten as: 
\begin{equation}
    \mathbf{z}=\mathbf{Y}\mathbf{a} \text{  and  } \mathbf{u}=\mathbf{X}\mathbf{a}.
\end{equation}
 
The DPD technique estimates the parameter $\mathbf{a}$ that minimizes the following cost function:
\begin{equation}
     \lVert \mathbf{e} \rVert^2 = (\mathbf{z} - \mathbf{u})^\top( \mathbf{z} - \mathbf{u})
\end{equation}
In practice, $\mathbf{a}$ is determined from a damped Newton algorithm as follows:
\begin{equation}
    \mathbf{a}_{p+1}=\mathbf{a}_p + \mu(\mathbf{Y}^\top\mathbf{Y})^{-1}\mathbf{Y}^\top\mathbf{e}.
\end{equation}
where $\mu$ is the relaxation constant.

The indirect learning method for estimating $\mathbf{a}$ requires the pseudo-inverse of $\mathbf{Y}$. 
Owing to the presence of low voltage values in some $y(n)$ at the rising and falling edges, $(\mathbf{Y}^\top\mathbf{Y})$ is typically ill-conditioned.
This results in a high condition number and increased susceptibility to numerical errors. 
Previous studies \cite{Chen19:Toward, Balatti21:Aircraft} addressed similar challenges by applying a low-rank approximation of $\mathbf{Y}$ using the TSVD technique \cite{Hansen87:truncatedSVD} to reduce the condition number, although these applications were not related to DPD. 

The matrix $\mathbf{Y}$ can be decomposed using SVD as follows:
\begin{equation}
\mathbf{Y}=\mathbf{U}\mathbf{S}\mathbf{V}^\top
 \end{equation}
where $\mathbf{Y}$ is an $L \times (KM+1)$ matrix; $\mathbf{U}$ is an $L \times L$ orthogonal matrix; $\mathbf{S}$ is an $ L \times (KM+1)$ diagonal matrix containing singular values sorted in descending order; and $\mathbf{V}^\top$ is a $(KM+1) \times (KM+1)$ orthogonal matrix.   
A low-rank approximation of $\mathbf{Y}$ can be constructed by removing the small singular values, along with the corresponding rows and columns in $\mathbf{U}$ and $\mathbf{V}$. 

By retaining only the largest $r$ singular values of $\mathbf{Y}$, the resulting low-rank matrix $\mathbf{{Y}}_r$ is expressed as follows:
\begin{equation}
\mathbf{Y}_r=\mathbf{U}_r\mathbf{S}_r\mathbf{{V}}^\top_r
\end{equation}
where $\mathbf{{U}}_r $ consists of the first $r$ columns of $\mathbf{U}$ and is an $L \times r$ matrix; $\mathbf{{S}}_r $ contains the first $r$ rows and columns of $\mathbf{{S}}$, making it an $r \times r$ matrix; and $\mathbf{V}^\top$ consists of the first $r$ rows of $\mathbf{{V}}^\top$, making it an $r \times (KM+1)$ matrix. 

The full SVD method utilizes all singular values of matrix $\mathbf{Y}$, including those that are extremely small.
These small singular values can amplify minor variations in the data, often arising from noise.
Consequently, the inclusion of small singular values can make the model more sensitive to noise and prone to overfitting.
In contrast, $\mathbf{Y}_r$ retains only the largest $r$ singular values, discarding the smaller ones.
This approach enhances robustness to noise and reduces susceptibility to overfitting.
The performance and sensitivity of $\mathbf{Y}_r$ for the DPD application are discussed in detail in the following section.

\section{RESULTS}
\label{sec:results}

\subsection{Testbed Setup}

\begin{figure*}
  \centering
  \includegraphics[width=0.75\linewidth]{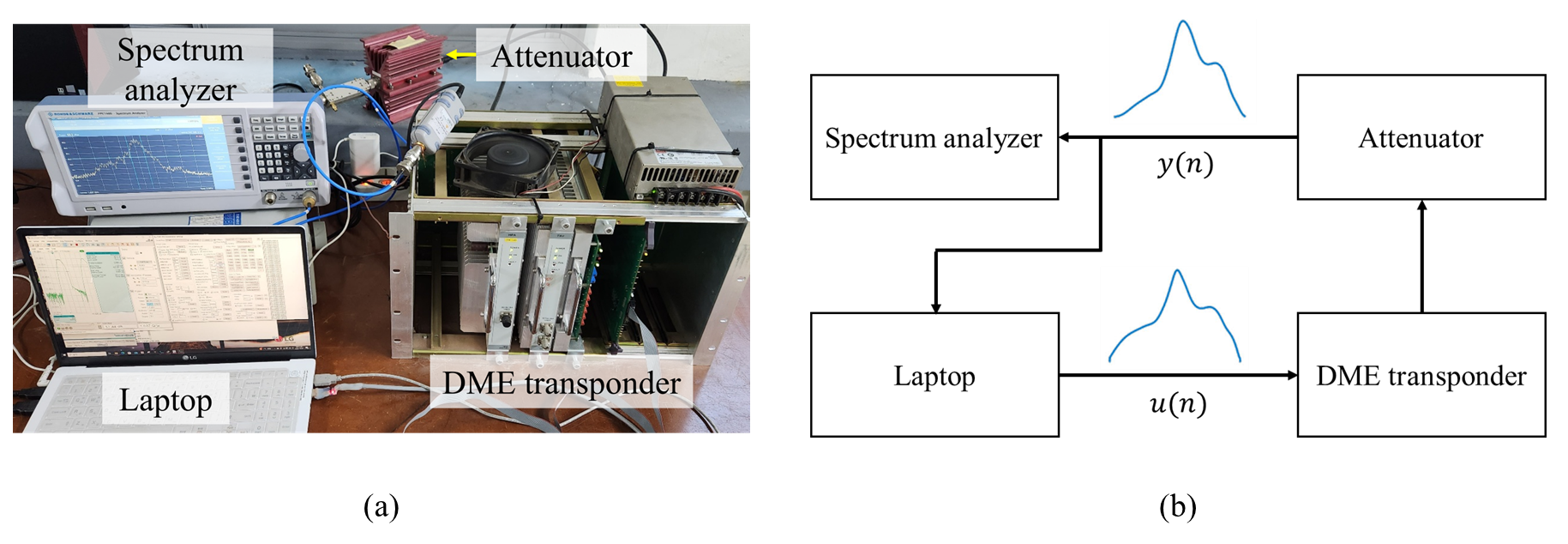}
  \caption{(a) Laboratory testbed setup and (b) corresponding block diagram.}
  \label{fig:testbed_combined}
\end{figure*}

A commercial 1000 W transponder (MOPIENS Inc.) was utilized for the DME transmitter testbed.
The channel and pulse repetition frequencies were set to 1037 MHz and 700 Hz, respectively.
A comprehensive view of the testbed setup is shown in Fig.~\ref{fig:testbed_combined}. 
In the testbed setup, the laptop generated a predistorted waveform, $u(n)$, using a predistorter based on a selected RF input digital pulse, $x(n)$.
The DME transponder received the predistorted pulse $u(n)$ and transmitted $y(n)$. 
Subsequently, the laptop received the transmitted pulse $y(n)$ and estimated the parameters for both the predistorter and postdistorter. 
To ensure compliance with DME power spectrum specifications, a spectrum analyzer measured the ERP of the transmitted pulse, $y(n)$. 
An attenuator was employed to reduce the power of the transmitted pulse, $y(n)$, ensuring safe signal reception by both the spectrum analyzer and the laptop.

In our experiment, the nonlinearity order ($K$) was selected within the range of 7–12, while the memory depth ($M$) was set to two. 
For each pair of $K$ and $M$ parameters, DPD was designed and tested for all ranks \(1 \leq r \leq KM+1\). The entire experiment was conducted using the SFOL24 pulse.

\subsection{Test Results}

The results of the tested parameter sets for $K$, $M$, and $r$ are listed in Table \ref{tab:Power_Spec_V}, showcasing those that satisfied the DME power spectrum specifications.
Notably, the transmitted pulse satisfied the power spectrum specifications for every nonlinearity order ($K$) only when $r = 12$.
In our DME transponder, an $r$ value of 12 for TSVD was found to be optimal for satisfying the DME power spectrum specifications.

\begin{table}
\centering
\caption{Comparison of SFOL24 power spectrum results with DME power spectrum specifications.}
\label{tab:Power_Spec_V}
\renewcommand{\arraystretch}{1.2}
\begin{tabular}{ccccc}
\hline
Adjacent channel (MHz) & $-2.0$     & $-0.8$    & $+0.8$    & $+2.0$      \\ \hline
Specifications (dBm)       & 3.0    & 23.0   & 23.0   &  3.0   \\ \hline
$K$=7, $M$=2, $r$=12 (dBm)        & $-0.3$    & 15.8   & 16.8   &  $-0.1$   \\
$K$=8, $M$=2, $r$=12 (dBm)        & 1.8    & 15.7   & 16.7   &  0.8   \\
$K$=9, $M$=2, $r$=12 (dBm)       & 0.8    & 16.3   & 17.0   &  0.4   \\
$K$=10, $M$=2, $r$=12 (dBm)        & 0.3    & 15.6   & 17.2   &  0.3   \\
$K$=11, $M$=2, $r$=12 (dBm)        & $-0.2$    & 16.2   & 18.1   &  $-0.7 $    \\
$K$=12, $M$=2, $r$=12 (dBm)    & 2.1    & 17.4   & 18.3   &  1.0      \\ \hline 
Theoretical pulse (dBm)        & $-11.3$    & 15.5   & 15.5   &  $-11.3$     \\
\hline
\end{tabular}
\end{table}

Table \ref{tab:Pulse_Shape_V} shows whether the pulse shape specifications are satisfied for the $K$, $M$, and $r$ parameter sets that met the DME power spectrum specifications outlined in Table \ref{tab:Power_Spec_V}. 
Notably, these parameter sets satisfied both the ICAO and FAA DME pulse shape specifications.

\begin{table}
\centering
\caption{Comparison of SFOL24 pulse shape results with ICAO and FAA DME pulse shape specifications.}
\label{tab:Pulse_Shape_V}
\renewcommand{\arraystretch}{1.2}
\resizebox{\columnwidth}{!}{
\begin{tabular}{cccc}
\hline
Pulse shape parameters   & Rise time ($\mathrm{\mu s}$)        & Width ($\mathrm{\mu s}$)          & Fall time ($\mathrm{\mu s}$)      \\ \hline
ICAO specifications & \textless 3.0  & 3.5($\pm0.5$)  & 3.0($+0.5$,$-0.5$)  \\
FAA specifications  & 2.5($+0.5$,$-1.0$)  & 3.5($\pm0.5$)  & 2.5($+0.5$,$-1.0$) \\
\hline 
$K$=7, $M$=2, $r$=12  & 2.80 & 3.15  & 2.96 \\
$K$=8, $M$=2, $r$=12  & 2.83 & 3.12  & 2.97 \\
$K$=9, $M$=2, $r$=12  & 2.79 & 3.10  & 2.94 \\
$K$=10, $M$=2, $r$=12 & 2.77 & 3.10  & 2.90 \\
$K$=11, $M$=2, $r$=12 & 2.75 & 3.10  & 2.89 \\
$K$=12, $M$=2, $r$=12 & 2.79 & 3.12  & 2.95 \\  \hline
Theoretical pulse     & 2.87 & 3.14 & 2.95 \\
\hline
\end{tabular}
}
\end{table}

\begin{figure}
  \centering
  \includegraphics[width=0.9\linewidth]{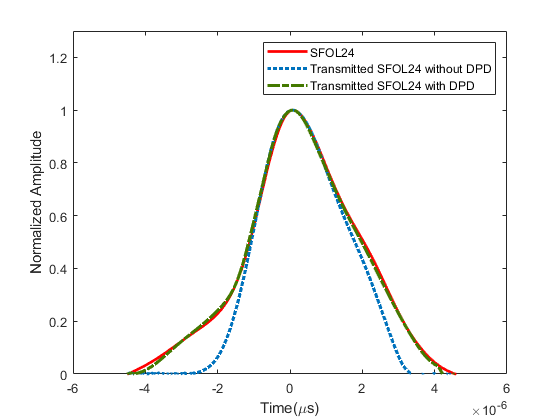}
  \caption{Comparison of the theoretical and transmitted SFOL24 pulses.}
  \label{fig:result_pulse}
\end{figure}

The transmitted pulses processed through the predistorter using the proposed DPD method and parameter sets successfully satisfied all DME specifications in our testbed. 
This highlights the effectiveness of the two strategies proposed in this study for transmitting SFOL pulses that satisfy DME specifications in high-power DME applications. 
Fig.~\ref{fig:result_pulse} shows that the transmitted SFOL24 pulses with the DPD method applied closely resemble the theoretical SFOL24 pulse. 
When the DPD method was not applied, the transmitted SFOL24 pulses displayed significantly different shapes compared to the theoretical SFOL24 pulse.

\begin{figure} 
  \centering
  \includegraphics[width=0.9\linewidth]{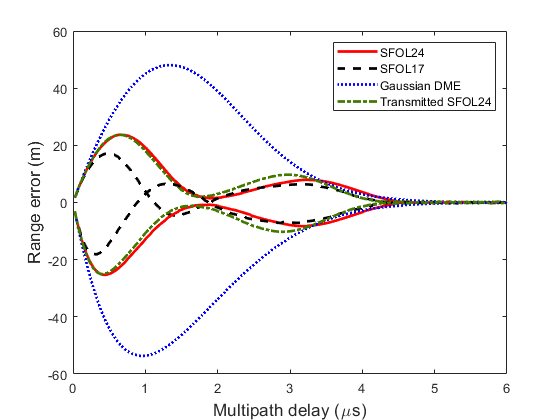}
  \caption{Comparison of the multipath-induced range errors of the transmitted SFOL24 pulse and the three theoretical pulses (Gaussian, SFOL17, and SFOL24) without actual transmission.}
  \label{fig:result_multipath}
\end{figure}

The multipath-induced range errors for the theoretical Gaussian, SFOL17, and SFOL24 pulses without actual transmission are presented in Fig.~\ref{fig:result_multipath} and Table~\ref{tab:Multipath_error}, along with the experimental results for the transmitted SFOL24 pulses. 
The findings suggest that the multipath-induced range errors were consistent across the proposed parameter sets and did not exhibit significant differences when compared with the theoretical pulse. 
In summary, the parameter sets that satisfied all DME specifications reduced the RMS of multipath-induced range errors by 62.8\% to 64.8\% compared with the Gaussian pulse.

\begin{table} 
\centering
\caption{Multipath-induced range errors of the transmitted SFOL24 pulses and the three theoretical pulses.}
\label{tab:Multipath_error}
\renewcommand{\arraystretch}{1.2}
\resizebox{\columnwidth}{!}{
\begin{tabular}{ccccc}
\hline
 \multicolumn{2}{c}{Multipath-induced range errors}  & In-phase (m) & Out-of-phase (m) & RMS (m) \\ \hline

SFOL24 & $K$=7, $M$=2, $r$=12         & 24.9              &  $-26.4$     &   9.7        \\
SFOL24 & $K$=8, $M$=2, $r$=12         & 24.4             &  $-26.0$     &   9.6        \\ 
SFOL24 & $K$=9, $M$=2, $r$=12         & 24.8             &  $-26.3$     &   9.6        \\ 
SFOL24 & $K$=10, $M$=2, $r$=12         & 24.7             &  $-26.1$     &   9.4        \\ 
SFOL24 & $K$=11, $M$=2, $r$=12         & 24.5             &  $-25.8$     &   9.3        \\ 
SFOL24 & $K$=12, $M$=2, $r$=12         & 23.7              &  $-25.1$     &   9.2        \\ 
SFOL24 & Theoretical pulse            & 23.8              &  $-25.3$     & 9.2          \\  \hline
Gaussian   & Theoretical pulse        & 48.1             &   $-53.7$  &    26.1         \\
\hline
SFOL17  & Theoretical pulse     & 17.1              & $-18.1$     & 6.0         \\
\hline
\end{tabular}
}
%\vspace{1mm} 
\flushleft
\noindent{\footnotesize Note: The table reports the worst-case (i.e., maximum absolute) multipath-induced range errors in both in-phase and out-of-phase conditions, along with the root-mean-square (RMS) of the total range errors.}
\end{table}

\section{CONCLUSION}
\label{sec:conclusion}

This study represents a significant advancement in the transmission of a variant SFOL pulse using a commercial Gaussian pulse-based DME transponder in 1000 W high-power mode. 
The SFOL17 pulse, while more robust to multipath effects compared to the Gaussian pulse, barely satisfies DME specifications. 
In a previous study, the DPD method enabled successful transmission of the SFOL pulse in 100 W low-power mode, meeting both pulse shape and ICAO power spectrum specifications. 
However, when applied to high-power DME, where noise levels are higher than in low-power DME, the transmitted SFOL17 pulse exceeded the power spectrum specifications.

To address this issue, two important strategies are proposed in this study. 
First, a variant SFOL (SFOL24) pulse was developed using a genetic algorithm, ensuring a sufficient margin for DME specifications while maintaining robustness against multipath effects. 
Second, the application of TSVD to the DPD technique was employed to ensure robustness against noise and prevent overfitting.

Experimental tests of these strategies on a Gaussian pulse-based commercial DME transponder demonstrated the successful transmission of the SFOL24 pulse, satisfying all DME specifications. 
The multipath-induced range errors of the transmitted SFOL24 pulses exhibited minimal deviations from the theoretical SFOL24 pulse, with a substantial 64.8\% reduction compared to the Gaussian pulse.

Therefore, the proposed strategies significantly enhance the ranging accuracy of DME by effectively mitigating multipath effects. 
These findings provide a foundation for the evolution of DME/DME navigation into not only a short-term APNT solution for aircraft but also a long-term APNT solution.

As future work, we plan to investigate practical methods for integrating SFOL pulse transmission and processing into existing DME avionics with minimal hardware modifications. 
We intend to make our implementation guidelines and experimental findings publicly available to support adoption by DME avionics manufacturers. 
We also anticipate that these advancements may inform regulatory updates and encourage civil aviation authorities to consider alternative pulse designs for higher-accuracy DME systems.

\bibliographystyle{./IEEEtaes}
\bibliography{./IEEEabrv,./mybibfile,./IUS_publications}

\end{document}